\documentclass[12pt]{article}
\begin{document}

\begin{center}
{\large \bf Information Measures for Inferring Quantum Mechanics}
\end{center}
\vspace{0.1in}

\begin{center}

{Rajesh R. Parwani\footnote{Email: parwani@nus.edu.sg}}

\vspace{0.3in}

{Department of Physics and\\}
{University Scholars Programme,\\}
{National University of Singapore,\\}
{Kent Ridge, Singapore.}

\vspace{0.3in}
\end{center}
\vspace{0.1in}
\begin{abstract}
Starting from the Hamilton-Jacobi equation describing a classical ensemble, one may infer a quantum dynamics 
using the principle of maximum uncertainty. That procedure requires an appropriate measure of uncertainty: Such a measure is constructed here from physically motivated constraints. It leads to  a unique single parameter extension of the classical dynamics that is equivalent to the usual linear quantum mechanics.

\end{abstract}

\vspace{0.5in}

\section{Deconstructing the Schrodinger equation}

Despite its remarkable quantitative success, quantum mechanics continues to puzzle us with its seemingly counter-intuitive predictions. Even the mathematical formalism most widely used for its description appears very different from that used in classical mechanics: one sees in quantum mechanics the appearance of complex numbers, probability amplitudes and an apparently exact linear evolution equation.

In this paper the structure of Schrodinger's equation, in particular its linearity,  will be derived within an information theoretic framework to be elaborated on below. The various assumptions involved in the derivation will also be discussed at length. 

Let us begin with a review of the Schrodinger equation for $N$ particles in $d+1$ dimensions, 
\begin{equation}
i\hbar \dot{\psi} = \left[ - {\hbar^2 \over 2} g_{ij} \partial_i \partial_j + V \right] \psi \, \label{schmulp}
\end{equation}
where $i,j= 1,2,......,dN$ and the metric is defined as $g_{ij} = \delta_{ij} /m_{(i)}$ with the symbol $(i)$ defined as   the smallest integer $\ge i/d$. That is, $i=1,...d$, refer to the coordinates of the first particle of mass $m_1$, $i=d+1,.....2d$, to those of the second particle of mass $m_2$ and so on. The overdot refers to a partial time derivative and the summation convention is used unless otherwise stated. Cartesian coordinates have been chosen as these allow an unambiguous correspondence between observables such as momenta and their operator representation \cite{Ballentine}.  

The metric $g_{ij}$ occurs naturally in the description of the system in configuration space \cite{synge,reg1} and plays a crucial role in the discussion below.
It is pertinent to note that the metric $g_{ij}$ is diagonal and positive-definite. This is a consequence of the form of the kinetic term in the Schrodinger equation in Cartesian coordinates. 

Since our intuition is mostly classical, it is useful to rewrite the Schrodinger equation in a form which allows comparison with Newtonian physics. The Madelung transformation \cite{mad} $\psi = \sqrt{p} \ e^{iS / \hbar}$ decomposes the Schrodinger equation into two real equations,
\begin{eqnarray}
\dot{S} + {g_{ij} \over 2} \partial_iS \partial_j S + V -{{\hbar}^2 \over 8} g_{ij} \left( {2 \partial_i \partial_j p \over p} - {\partial_i p \partial_j p \over p^2} \right) &=& 0 \, ,  \label{hj3} \\
\dot{p}  + g_{ij} \ \partial_i \left( p \partial_j S \right)  &=& 0 \, .  \label{cont3} 
\end{eqnarray}
The first equation is a generalisation of the usual Hamilton-Jacobi equation, the term with explicit $\hbar$ dependence (the ``quantum potential" \cite{Ballentine}) summarising the peculiar and nonlocal aspects of quantum theory. The second equation is a 
continuity equation expressing the conservation of probability, $ \int p(x,t) dx^{Nd}$.

Equations (\ref{hj3}, \ref{cont3}) may be obtained from a variational principle \cite{reg1}, one minimises the action
\begin{eqnarray}
\Phi &=& \int  p \left[ \dot{S}  + {g_{ij} \over 2} \partial_i S \partial_j S   + V \right] dx^{Nd}  dt   \ + {{\hbar}^2 \over 8}   I_F \, \label{varmulp} 
\end{eqnarray}
with respect to the variables $p$ and $S$. Interestingly, the quantity
\begin{eqnarray}
I_F & \equiv & \int dx^{Nd}  dt \  g_{ij} \ p \ (\partial_i \log p) (\partial_j \log p) \,  \label{fish}
\end{eqnarray}
resembles the ``Fisher information" \cite{fisher}, whose inverse sets a lower bound on the variance of the probability distribition $p(x)$ through the Cramer-Rao inequality \cite{Kullback,reg1}. Since a broader probability distribution $p(x)$ represents a greater uncertainty in $x$, the term $I_F$ is actually an inverse uncertainty measure.

The equations (\ref{varmulp}, \ref{fish}) were used in Ref.\cite{reg1,fried1} to derive Schrodinger's equation through a procedure analogous to the principle of maximum entropy (uncertainty) \cite{jay,Kapur} used in statistical inferrence theory. The idea is that without the term $I_F$, variation of Eq.(\ref{varmulp}) gives rise to equations describing a classical ensemble. As the probability distribution $p(x)$ characterising the ensemble is supposed to represent some fluctuations of unknown origin, we would like to be as unbiased as possible in its choice. This is achieved by choosing the broadest distribution possible, representing our maximum uncertainty. Technically, this is implemented in (\ref{varmulp}) by minimising (\ref{fish}) when varying the classical action: $ {\hbar}^2 /8 $ is the Lagrange multiplier.

\section{Constructing the Measure}
However, the intriguing approach of Ref.\cite{reg1} does not explain, {\it a priori}, the form of the information measure that should be used. That is, why must $I_F$ be minimised rather than something else?

The goal of this section is to construct, from first principles, information measures that are permissible. To fix the notation, consider therefore the same classical ensemble as in Sec.(1), but now constrained by a general information measure $I$ whose form is to be determined. The relevant action is then

\begin{eqnarray}
{\cal{A}} &=& \int  p \left[ \dot{S}  + {g_{ij} \over 2} \partial_i S \partial_j S   + V \right] dx^{Nd}  dt   \ + \lambda   I \, , \label{gen} 
\end{eqnarray}
with $\lambda$ a Lagrange multiplier. Varying this action will give rise, in general, to a nonlinear Schrodinger equation after an inverse Madelung transformation, 
\begin{equation}
i\hbar \dot{\psi} = \left[ - {\hbar^2 \over 2} g_{ij} \partial_i \partial_j + V \right] \psi \, + \ F(\psi,\psi^{\dag}) \psi \, , \label{nonsch}
\end{equation}
with $F$ representing the nonlinearity.

In order to construct an explicit form for the information measure $I$, constraints need to be imposed. These constraints are of two types. The first type are those that are required for $I$ to  be sensibly interpreted as a measure of inverse uncertainty (information). The condition $[S1]$ below belongs to this type.

The other constraints $[S2]-[S6]$ to be discussed below are of a different type. One may adopt two alternative perspectives to motivate these constraints. The first, classical, perspective is to view the action (\ref{gen}) as  a generalised form of classical dynamics. In that case a minimal deformation of the usual classical dynamics is achieved if the additional constraints are the same as those already satisfied by the
$I=0$ part of (\ref{gen}): locality, homogeneity, separability, Gallilean invariance, and absence of more than two derivatives in each product of terms in the action. Thus by using $[S2]-[S6]$ one is not imposing any constraints on the action that do not already exist. In this classical perspective the physical motivations for the constraints are either the usual ones with an obvious interpretation (locality, separability, Gallilean invariance), or else they are explained below.     

The second, quantum, perspective is to view the conditions $[S2]-[S6]$ with respect to the interpretation of Eq.(\ref{nonsch}), which represents a generalised Schrodinger equation. As the usual linear quantum mechanics has been experimentally well tested, one could argue that it makes sense to only consider those potential deformations of the linear theory  that permit as much of the usual interpretations of the wave-function as possible. Remarkably, as discussed below, the same conditions $[S2]-[S6]$ motivated by the classical perspective are needed also for this quantum viewpoint.

Here then are the axioms:

\begin{itemize}
\item $[S1]$ (i) Firstly, by definition, the measure $I$ should be a real-valued and positive definite functional for 
{\it all} $p= \psi^{\dag} \psi$. More specifically, we would like the measure to be universal in the sense of being independent of the external potential $V.$ \\
$\;\;\;\;\;\;$ (ii) Also, the interpretation of $I$ as an information (inverse uncertainty) measure requires that it should approach a minimum when $p$ is uniformly distributed. (A minimum exists because by (i) $I$ is positive definite).

\item $[S2] \;$ Locality: $I$ should be of the form $I = \int dx^{Nd} dt \ p  H(p)  $ where $H$ is a function of the probability $p(x,t)$ and and its spatial derivatives.  In the quantum perspective this local form ensures the validity of the weak superposition principle in the equations of motion (\ref{nonsch}): states with negligible overlap will not influence each other strongly. 
(In principle one should also allow $I$ to depend on $S, x$ and time derivatives. These generalisations are discussed in the next section.) 

\item $[S3] \;$ Homogeneity: $H$ should be invariant under scaling, $H(\lambda p) = H(p)$. The normalisation of probability, $1 = \int dx^{Nd} p(x,t)$, implies that the dimensions of $p(x,t)$ depend on the dimensions of the configuration space. Thus by demanding $H$ to be scale invariant one ensures that the resulting equations of motion, whether in the classical or quantum perspective, have a universal form independent of the number of particles. Thus this condition may be viewed as restricting the search to universal dynamics.\\
(There is another motivation sometimes given for this homogeniety condition: It  allows solutions of (\ref{nonsch}) to be (re)normalised and thus allow for the usual interpretation of states after a measurement process, as discussed in \cite{kibble}.) 

\item $[S4] \;$ Separability: $H$ should be separable for the case of two independent sub-systems described by probability distributions $p_1$ and $p_2$: $H(p=p_1 p_2) =H(p_1) + H(p_2)$.

\item $[S5] \;$ $H$ should be Gallilean invariant\footnote{In this paper, when discussing the symmetry of the Schrodinger equation one will always refer to the case of vanishing potential, $V=0$.}.

\item $[S6] \;$ $H$ should not contain more than two derivatives in any product of terms that appears in it. As each derivative involves an inverse length, this condition obviously restricts the number of new dimensional parameters that can appear in (\ref{gen},\ref{nonsch}). This condition will be referred to in brief as ``absence of higher number of derivatives" or ``AHD". \\
As will be apparent later, the implementation of this condition means that the lagrange multiplier $\lambda$ in (\ref{gen}), and hence Planck's constant, is the only new parameter that is required in making the transition from classical to quantum mechanics. Conversely, relaxing the AHD condition would imply the appearance of other parameters, with the dimensions of length, and a generalised form of quantum dynamics.  

\end{itemize}

I would like to reiterate that the conditions $[S2]-[S6]$  above are all already satisfied by the {\it classical} part of the action (\ref{gen}), so demanding them of the additional piece $I$ is actually quite natural and minimalist. The additional motivations provided by viewing the equations from the quantum perspective are simply a bonus. 

Having enumerated and motivated the axioms, one may begin constructing the measure. 
Clearly the homogeneity requirement $[S3]$ cannot be satisfied if $H$ depends only on $p$ : it must also contain   derivatives of $p$. The AHD condition and rotational invariance imply that the building blocks of $H$ must be 
\begin{equation}
g_{ij} U_1 \partial_i U_2 \partial_j U_3 \;\; \mbox{and} \;\; g_{ij} V_1 \partial_i \partial_j V_2 \,, \label{bb}
\end{equation}
where the $U_i, V_i$ are functions of $p$.  
Separability can now be used to deduce that $H$ must be linear in $g_{ij}$: 
\begin{equation}
H = g_{ij} \left( U_1 \partial_i U_2 \partial_j U_3 + V_1 \partial_i \partial_j V_2 \right) \, . \label{ht}
\end{equation}
One may consider sums of such structures and so place an additional index $n$ on the $U_i, V_i$ but it is easy to check that the final result below remains unchanged.  One could also use separability to restrict the explicit forms that $U,V$ may take but I will use the homogeneity condition for that purpose below. (Note that if the form of $H(p)$ were restricted so that derivatives of $p$ occured only in polynomial form, as is commonly done in physics, then separability would not be required in obtaining (\ref{ht})).

Using the chain rule one can rewrite (\ref{ht}) as 
\begin{equation}
H= g_{ij} \left(  {\partial_i p \partial_j p \over p^2} [U_1 U_{2}^{\prime} U_{3}^{\prime} p^2 + V_{1} V_{2}^{\prime \prime} p^2] + {\partial_i \partial_j p \over p} [V_1 V_{2}^{\prime} p ] \right) \, , \label{ht2}
\end{equation}
where the prime symbol denotes a derivative with respect to $p$. Consider now the scaling $p \to \lambda p$ under which $H$ is required to be invariant. The terms inside the square  brackets become dependent only on the product $\lambda p$. Since the forms  $\partial_i p \partial_j p \over p^2$ and $\partial_i \partial_j p \over p$ are distinct and independently scale invariant, this means that the terms in square brackets must also be independent of $\lambda$: but since those terms depend only on $\lambda p$, this implies that the terms in square brackets are simply constants. 

Thus one obtains
\begin{equation}
I= \int dt dx^{Nd} p g_{ij} \left( A{\partial_i p \partial_j p \over p^2}  +  B {\partial_i \partial_j p \over p} \right) \, . \label{ht2}
\end{equation}
The ``$B$" term  gives a surface contribution which might not vanish for some wavefunctions and so its contribution to $I$ is of indefinite sign. The positivity and universality of $I$ therefore requires us to choose $B=0$.

Hence one concludes that the 
unique solution of the conditions $[S1]-[S6]$ is the information measure $I_F$ given in (\ref{fish}).  The Lagrange multiplier $\lambda$ in (\ref{gen}) must then have the dimension of (action)$^2$ thereby introducing the Planck constant into the picture; the equation of motion is then the {\it linear} Schrodinger equation. 

Interestingly, neither the second part of  $[S1]$ nor the full Gallilean invariance was used explicitly in the above construction eventhough  the final result, the  measure $I_F$, does satisfy all the conditions. However these additional constraints will be useful in the next section.

It should be noted that, as shown in \cite{reg2}, the action (\ref{varmulp}) increases for variations of $p$ that keep $S$ fixed, the increase being due to an increase in $I_F$, so that the resulting solutions are not just an extremum but do indeed minimise the information measure $I_F$. 

The positivity condition in $[S1]$ plays an important physical role beyond ensuring the existence of a minimum for $I$ (state of maximum uncertainty). It also guarantees  that the following energy functional is bounded from below for potentials $V$ that are likewise bounded:
\begin{eqnarray}
{\cal{E}}[S,p] &=& \int d^{Nd} x \ p \left( {g_{ij} \over 2} \partial_i S \partial_j S + V + \lambda H(p)   \right)  
\end{eqnarray}
where the function $H$ is defined in $[S2]$. This functional is conserved for stationary states and it also reduces to the average of the usual quantum mechanical hamiltonian for the case of the linear theory. These properties of the energy functional qualify it as the most natural to use for defining the energy of the system in a potential generalisation (nonlinear) of quantum theory.

The meaning of the AHD condition can be elucidated with an explicit example. Consider
\begin{equation}
H_1(p) = g_{ij} \partial_i (\log p + \eta f(p)) \partial_j (\log p + \eta f(p)) 
\end{equation}
with $ f(p)= g_{kl} (\partial_k \log p) (\partial_l \log p)$ and $\eta$ a constant. This $H_1$ satisfies all the constraints except AHD. The price to pay is the appearance of an additional parameter, $\eta$, required to balance the dimensions of the higher order derivatives. 

Thus the AHD condition ensures that, within the information theoretic approach, the Schrodinger equation is the unique {\it single} parameter extension of the classical statistical Hamilton-Jacobi equations. Since the information theoretic approach attempts to provide an unbiased description of data, one may say that the AHD condition further restricts us to the simplest unbiased description.

\section{Relaxing some conditions}
It has been implicitly assumed in the last section that the  metric $g_{ij}$ that appears in $I$ is the same as the metric in the classical part of the action. If one allows in the measure a  metric $\bar{g}_{ij}$ which is still diagonal but different from the classical metric $g_{ij}$ then a nonlinear Schrodinger equation apparently ensues. However the nonlinearity can be removed by a change of variables (a nonlinear gauge transformation) \cite{doeb,aub} with the result that for a range of values of the Lagrange multiplier $\lambda$ one actually recovers the usual linear Schrodinger equation. (This example highlights the point that a nonlinear Schrodinger equation cannot immediately, and in all generality, be declared pathological). However for the remaining values of $\lambda$ one obtains after the change of variables \cite{aub} a linear {\it diffusion} equation. It will be assumed for the rest of this paper that the classical metric is used also in the information measure when symmetries are preserved.

Consider now allowing $I$ and hence $H$ to depend also on the phase factor $S(x,t)$ in addition to the probability density $p(x,t)$. By definition, the phase factors $S_1$ and $S_2$ for two independent systems are additive in the composite system, 
$S= S_1 + S_2$. One can proceed as before and consider the most general structures restricted by rotational invariance, AHD, homogeneity, separability and positivity. The result is a generalised measure of the form    
\begin{equation}
I_Q = \int dt dx^{Nd} \ p g_{ij} \partial_i( \alpha_1 \log p + \alpha_2 S)  \partial_j (\alpha_1 \log p + \alpha_2 S)   , \label{I2}
\end{equation}   
where the $\alpha_k$ are constants. In arriving at this structure, homogeneity has only been used to imply that derivatives of $p$ occur, while separability is the stronger constraint as it acts also on the $S$ variable. 
However the second part of the condition $[S1]$  requires that $\alpha_2 =0$, thus eventually one is again led to the Fisher form. (Again, more generally one may place another index $n$ on the $\alpha_k$ and sum over such terms but the conclusion remains unchanged.) 

Nevertheless, the special case of (\ref{I2}) with $\alpha_1=0$ but $\alpha_2 \neq 0$ is sufficiently interesting to deserve further study because like the classical measure $I_F$, it is positive definite by itself, but unlike $I_F$ it also contains some information about the phase of the wavefunction. Now, if used in the variational action, this $S$ dependent term can be absorbed, after a scaling of the metric, by a similar term already existing in the classical part of the action (\ref{gen}). The net result is therefore still a linear Schrodinger equation but with a mass, $\bar{m}_{(i)}$  which is renormalised with respect to the original mass parameter $m_{(i)}$ in the classical theory. Empirically this renormalisation will have no consequence if all calculations, as usual,  refer to the mass parameter appearing in the quantum theory.

One may also consider allowing time derivatives of $p$ and $S$ in $H$. However the demands of positivity conflict with those of separability unless one relaxes the AHD condition: then it is possible to have structures such as 
\begin{equation}
H_2(p) = g_{ij} \partial_i \left(\log p + \eta {\dot{p} \over p} \right) \partial_j \left(\log p + \eta {\dot{p} \over p} \right) 
\end{equation}
that contain dimensionfull parameters.

Finally, one may consider an explicit dependence on the coordinates, $x_i$, in $H$. However such terms are ruled out by translational invariance.

\section{Conclusion}
If one adopts the philosophy that the laws of physics should be constructed so as to provide the most economical and unbiased representation of empirical facts, then the principle of maximum uncertainty \cite{jay, Kapur} is the natural avenue by which to investigate the foundations of quantum theory \cite{fried1,reg1}. 

The investigation here has extended the initiative of \cite{reg1} in several ways. Firstly, the constraints that a relevant  information measure should satisfy have been made explicit and motivated from two alternative perspectives. Secondly, the measure has been constructed from the  constraints rather than postulated, thus motivating the structure of the linear Schrodinger equation. Indeed, it has been shown here that within the information theoretic approach, the linear Schrodinger equation is the unique one-parameter extension of the classical dynamics.

One should compare the approach here with an alternative, axiomatic, but not information theoretic based, construction of the Schrodinger equation from classical mechanics  discussed in \cite{HR}. Starting from the classical Hamilton-Jacobi equations, the authors add constrained fluctuations to the kinetic term. The result is an equation similar to Eq.(\ref{gen}) above with an explicitly postive definite $H(p)$ and with the symmetries of the classical action. However, in addition to the major differences in motivation and epistemology, there is a significant technical contrast between this paper and Ref.\cite{HR}: there the authors postulate and use an ``exact uncertainty relation" instead of the homogeneity condition $[S3]$ adopted here. As discussed above, the homogeneity condition is already satisfied by the classical action and has the interpretation of making the form of the equations of motion  independent of the dimension of configuration space, and thus it is a simple restriction to universal dynamics. 

An open and interesting problem is to extend the constructive approach adopted in Sec.(2) to include spin \cite{reg3} and relativistic effects \cite{reg1}. This might involve further refinement of the conditions in Sec.(2) and might result in the use of more general information measures such as $I_{Q}$ of eq.(\ref{I2}).

Finally, one may enquire into the possible generalisations of quantum mechanics (typically non-linear) that result from omitting one or more of the conditions $[S1]-[S6]$ and the ensuing phenomenological consequences. These issues are discussed elsewhere \cite{raj1,raj3}. It is of interest to note that one of the earliest suggestions for a nonlinear Schrodinger equation \cite{BM} had a $\log p$ nonlinearity, which is allowed if the homogeneity condition is abandoned. \\

\noindent{\bf Acknowledgement}\\
I am grateful to Michael Hall and Marcel Reginatto for critical readings of earlier drafts of this paper and for many stimulating discussions.

\end{document}